# Has the Brain Maximized its Information Storage Capacity?

**Armen Stepanyants**
Cold Spring Harbor Laboratory
1 Bungtown Rd., Cold Spring Harbor, NY 11724
*stepanya@cshl.edu*

## Abstract

Learning and memory may rely on the ability of neuronal circuits to reorganize by dendritic spine remodeling. We have looked for geometrical parameters of cortical circuits, which maximize information storage capacity associated with this mechanism. In particular, we calculated optimal volume fractions of various neuropil components. The optimal axonal and dendritic volume fractions are not significantly different from anatomical measurements in the mouse and rat neocortex, and the rat hippocampus. This has led us to propose that the maximization of information storage capacity associated with dendritic spine remodeling may have been an important driving force in the evolution of the cortex.

## Introduction

Many important brain functions, such as learning and memory, depend on the plasticity of neuronal circuits. Traditionally, plasticity is thought to rely on the following biological mechanisms: changes in the strengths of existing synaptic connections, formation and elimination of synapses without remodeling of neuronal arbors, and remodeling of dendritic and axonal branches. In order to understand the respective roles of these mechanisms we began to evaluate their plasticity potentials, i.e. the number of available yet different circuits attainable by each mechanism. In particular, we calculated the plasticity potential associated with the reorganization of neuronal circuits by dendritic spine remodeling. We expressed our results in terms of the logarithm of the number of available circuits, or the information storage capacity.

Next, we looked for geometrical parameters of the cortical circuits, which maximize information storage capacity associated with this mechanism. We found optimal axonal, and dendritic volume fractions as functions of dendritic and axonal length densities, average dendritic spine length, and density of synapses. We compared these optimal geometrical parameters with the anatomical data and found a reasonable agreement.

## Information storage capacity

We start by briefly reviewing the framework behind the calculation of the information storage capacity due to the formation and elimination of synapses, which has been done previously[1,2]. Because the majority of excitatory synapses are located on spines[3] the reorganization of neuronal circuits can be implemented by retracting some dendritic spines from pre-synaptic axons and extending them towards other axons. Such switching of pre-synaptic partners is only possible if there are a greater number of axons within a spine length of a dendrite than the number of dendritic spines (Fig. 1). We refer to locations in neuropil where an axon is present within a spine length of a dendrite as potential synapses. Potential synapse is a necessary but not sufficient condition of an actual synaptic connection.

If the number of potential synapses is equal to the number of spines (Fig. 1B), then there are no available pre-synaptic partners and spine remodeling cannot contribute to circuit reorganization. If the number of potential synapses is greater than the number of spines (Fig. 1C), then spine remodeling can contribute to circuit reorganization (Fig. 1D).

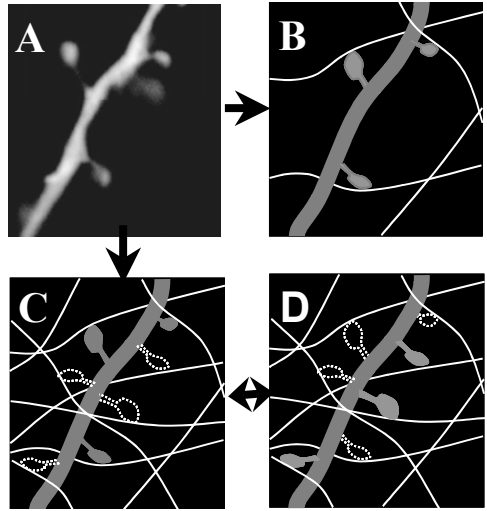

Figure 1: Spine remodeling as a mechanism of circuit reorganization[1,2]. **A.** Spiny dendrite in macaque neocortex visualized by light microscopy. **B**. A sketch showing the dendritic branch from A together with adjacent axons (thin lines). The number of potential synapses is equal to the number of spines and spine remodeling cannot contribute to circuit re-organization. **C**. An alternative scenario in which spine remodeling can contribute to circuit re-organization. Actual dendritic spines (solid gray) for actual synapses. Potential synapses include both actual synapses and other possible spine locations (dotted contours). The number of potential synapses is much greater than the number of spines. **D**. New circuit obtained from **C** through spine remodeling.

To determine which scenario (Fig. 1B or Fig. 1C) better reflects the real brain, we have derived a mathematical expression (making no assumptions about neuronal arbor shapes) to evaluate the ratio of numbers of actual and potential synapses. We call this ratio the filling fraction, $f$:

$$f = \frac{n_s}{\overline{\sin(\theta)} 2 s \rho_a \rho_d} \quad (1)$$

Here $s$ is the spine length (measured from the tip of the spine to the midline of the dendritic branch), $\rho_a$ and $\rho_d$ are axonal and dendritic length densities, and $\overline{\sin(\theta)}$ is the mean sine of the angle between axonal and dendritic branches forming potential synapses. This mean sine is equal to $\pi/4$ for uniformly distributed axonal or dendritic branches (cortex, hippocampus), and to $1$ for axons and dendrites intersecting at the right angle (parallel fibers of granular cells and dendrites of Purkinje neurons in cerebellum). The filling fraction calculated according to Eq. 1 for different species and brain areas[1,2] is in the range of 0.1-0.3. As a result, cortical micro-architecture is depicted better in Figs. 1C,D rather than in Fig. 1B.

The filling fraction, $f$, is a measure of plasticity potential associated with spine remodeling, as it reflects the number of different circuits that can be realized in given neuropil volume through spine reorganization. The information storage capacity of unit volume of neuropil due to spine remodeling is defined as base two logarithm of the number of different synaptic connectivity patterns. We have shown that the information storage capacity depends on the filling fraction as[1]:

$$i = n_s \left(1.25 - \log_2 f\right). \qquad (2)$$

Eq. (2) is an approximation of a more general expression for information storage capacity[1,2], made for biologically relevant values of the filling fraction, $f < 0.4$.

## Optimal neuropil

A unit volume of neuropil can be broken down into several components,

$$\kappa_a + \kappa_d + \kappa_{syn} + \kappa_{rest} = 1, \qquad (3)$$

where $\kappa_a$ and $\kappa_d$ are the axonal (not including boutons) and dendritic (without spines) volume fractions. $\kappa_{syn}$ is the synaptic volume fraction, which includes dendritic spines, axonal boutons, and the part of glia dedicated to synapse maintenance. $\kappa_{rest}$ denotes the remaining part of the neuropil, including the rest of glia and the extracellular space.

At first glance, it might appear that in order to obtain large structural synaptic information storage capacity, $i$, it is sufficient to increase the axonal and/or dendritic length densities, $\rho_{a,d}$. This change would result in the decrease of the filling fraction, $f$ [according to Eq. (1)], and consequently, increase in the information capacity, $i$, Eq. 2. However, due to the constraint on the volume of neuropil, Eq. (3), the increase in $\rho_{a,d}$ will be accompanied by a reduction in the synaptic density, $n_s$. This reduction, in turn, will have an opposite effect on information capacity. The three considered components, i.e. axonal, dendritic, and synaptic, have to be perfectly balanced in order to achieve the maximum information capacity. This balance is realized when axons and dendrites occupy equal fractions of neuropil given by the following function of the filling fraction only (see Methods for details):

$$\kappa_a = \kappa_d = \frac{1 - \kappa_{rest}}{1.87 - \ln f}. \qquad (4)$$

This function of the filling fraction is illustrated in Fig. 2 for a special case of $\kappa_{rest} = 0$.

## Anatomical data

Axonal and dendritic volume fractions obtained with electron microscopy and corresponding filling fractions are presented in Table 1. The filling fraction for the CA1 field of the rat hippocampus is based on CA3 to CA1 projection only[1,2].

|  | Axonal volume fraction $\kappa_a$ | Dendritic volume fraction $\kappa_d$ | Filling fraction $f$ |
|---|---|---|---|
| Mouse and rat neocortex | $0.31 \pm 0.09^4$ $0.36 \pm 0.03^5$ $0.34^3$ | $0.24 \pm 0.07^4$ $0.23 \pm 0.02^5$ $0.35^3$ | $0.26^{1,2}$ |
| Rat hippocampus CA1 (CA3 to CA1 projection) | $0.29 \pm 0.03^5$ | $0.26 \pm 0.03^5$ | $0.22^{1,2}$ |

Table 1: Axonal, dendritic volume fractions, and corresponding filling fractions for the mouse and rat neocortx, and the rat hippocampus.

In comparing these volume fractions with the optimal fractions, Eq. (4), we set $\kappa_{rest} = 0$. This is justified by the fact that $\kappa_{rest}$, which consists mainly of the extracellular space, is significantly reduced in serial section electron microscopy preparation. The anatomical fractions from Table 1 contain a large spread, and are within 20% of the theoretically predicted value (Fig. 2). A more consistent study, which would measure volume fractions and filling fractions from the same brain, is needed in order to further justify the hypotheses that neuropil is optimally designed to store information in patterns of synaptic connectivity.

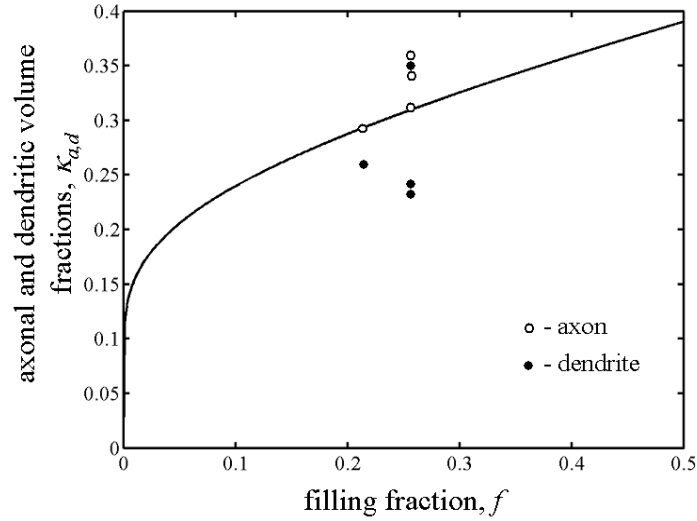

Figure 2: Optimal volume fractions of axons and dendrites as a function of the filling fraction, $f$, solid line. Circles represent anatomical data from Table 1.

## Conclusion

We calculated volume fractions occupied by axons and dendrites in a neuropil optimally designed to maximize information storage capacity due to spine remodeling. These optimal volume fractions are not significantly different from those measured anatomically. This leads us to suggest that maximizing information storage capacity due to spine remodeling may have been an important driving force in the evolution of the cerebral cortex.

## Methods

In this section we derive the expression for the optimal values of axonal and dendritic volume fractions [Eq. (4) in the main text]. This optimization problem, constrained by the fact that neuropil consists of several particular components, Eq. (3), is equivalent to the problem of finding the maximum of the Lagrange function,

$$I = i - \lambda(\kappa_a + \kappa_d + \kappa_{syn} + \kappa_{rest}), \tag{5}$$

where $\lambda$ is a positive parameter.

We search for the maximum of the function $I$ by varying axonal and dendritic length densities, $\rho_{a,d}$, and density of synapses, $n_s$, while keeping basic neuropil properties, i.e. basic geometry of axonal and dendritic arbors, geometrical factor $\overline{\sin(\theta)}$ and the average spine length, $s$, unchanged. Since, axonal and dendritic volume fractions, $\kappa_{a,d}$, and their length densities, $\rho_{a,d}$, scale with total length of axons and dendrites per neuron respectively, and the synaptic volume fraction, $\kappa_{syn}$, is proportional to the density of synapses, $n_s$, we have,

$$\frac{\partial \kappa_{a,d}}{\partial \rho_{a,d}} = \frac{\kappa_{a,d}}{\rho_{a,d}}, \quad \frac{\partial \kappa_{syn}}{\partial n_s} = \frac{\kappa_{syn}}{n_s}. \tag{6}$$

To find the optimal volume fractions of neuropil components we first set the partial derivatives of the function $I$ with respect to $\rho_{a,d}$ and $n_s$, to zero,

$$\frac{\partial I}{\partial \rho_{a,d}} = \frac{n_s}{\rho_{a,d} \ln 2} - \lambda \frac{\kappa_{a,d}}{\rho_{a,d}} = 0$$

$$\frac{\partial I}{\partial n_s} = 1.25 - \log_2 f - \frac{1}{\ln 2} - \lambda \frac{\kappa_{syn}}{n_s} = 0 \tag{7}$$

Next, we exclude parameter $\lambda$ from these equations and receive the relations among optimal synaptic, axonal, and dendritic volume fractions,

$$\kappa_{syn} = -\kappa_{a,d}(0.13 + \ln f). \tag{8}$$

Finally, by substituting synaptic volume fraction, $\kappa_{syn}$, from Eq. (3) into Eq. (8) we arrive at:

$$\kappa_a = \kappa_d = \frac{1-\kappa_{rest}}{1.87 - \ln f} \; . \tag{9}$$

This is Eq. (4) of the main text.

## Acknowledgments

I thank Dr. D.B. Chklovskii for numerous discussions and support during the writing of this manuscript.